%Paper: astro-ph/9405021
%From: pyne@cfa160.harvard.edu (Ted Pyne)
%Date: Tue, 10 May 94 12:10:23 EDT

%%%%%%%%%%%%%%%%%%%%%%%%%%%%%%%%%%%%%%%%%%%%%%%%%%%%%%%%%%%%%%%%%%%%%
% This paper comes with one figure appended at the end (after \bye) %
% as a postscript file. The paper will tex normally if you ignore   %
% the figure.                                                       %
%%%%%%%%%%%%%%%%%%%%%%%%%%%%%%%%%%%%%%%%%%%%%%%%%%%%%%%%%%%%%%%%%%%%%

\magnification=1200
\vbadness=10000
\hfuzz=10pt \overfullrule=0pt
\baselineskip=12pt
\parindent 20pt \parskip 6pt

%\line{\hfil DRAFT: apjlnet, 5/9/94}
%\vskip1.0truein
\centerline{\bf Beyond the Thin Lens Approximation}
\vskip0.25truein
\centerline{Ted Pyne and Mark Birkinshaw}
\centerline{Harvard-Smithsonian Center for Astrophysics}
\centerline{60 Garden St., Cambridge, MA 02138}
\vskip0.25truein
\centerline{Submitted to ApJl: 9 May 1994}
\vskip0.5truein

\noindent
{\bf Abstract}

We obtain analytic formulae for the null geodesics of
Friedmann-Lema\^{\i}tre-Robertson-Walker
spacetimes with scalar perturbations in the longitudinal gauge.
We use these to provide
a rigorous derivation of the cosmological lens equation.
We obtain an expression for the magnification
of a bundle of light rays in these spacetimes
without restriction to static or thin lens
scenarios. We show how
the usual magnification matrix
naturally emerges in
the appropriate limits.


\noindent
{\bf 1. Introduction}

The bending of light by a single symmetric gravitational
lens in a Euclidean space is shown in Fig. 1. The symmetry
guarantees that the lines of sight from the observer, $o$,
to the lens, $l$, and to both the lensed, and
unlensed, image of the emitter, $e$, lie in the same plane,
as do the angles $\alpha$, $\beta$, and $\theta$.
$D(o,l)$ and $D(o,e)$ are the distances from the
observer to the lens and emitter, respectively.
We assume that the deflection angle, $\alpha$, is small.
Then,
locally about the line of sight to the emitter's image,
we may approximate the two-spheres at distances
$D(o,l)$  and $D(o,e)$ from the observer as planes, called the
lens plane and source plane respectively.
$D(l,e)$ denotes the distance between these two planes.
The assumption of small deflection angle also allows us to relate the
lensing angles by $\beta =\theta -\alpha D(l,e)/D(o,e)$.
This is the simplest example of the gravitational lens equation.

The generalization of this equation to more complicated
lens structures and non-Euclidean background spaces
proceeds by a number of steps. A general lens is not
symmetric so that the angles $\alpha$,
$\beta$, and $\theta$ are not necessarily coplanar. To handle this,
consider a set of cartesian
axes with origin at the observer. Choose the $x$-axis to
coincide with the line-of-sight to the image. Let $\alpha^i$, where
$i$ runs over $\lbrace 2,3\rbrace$,
be the angle between the $x$-axis and the projection
into the $x^ix$-plane of the line-of-sight vector
from the deflection point, $p$, to the emitter. Similarly, let
$\beta^i$ be the angle between the
projection into the $x^ix$-plane of the line-of-sight
vector from the observer to the lens and the
projection into the $x^ix$-plane of the line-of-sight
vector from the observer to the unlensed image.
Also, let $\theta^i$ be the angle between the $x$-axis
and the projection into the
$x^ix$-plane of the line-of-sight vector from the
observer to the lens.  To allow for non-Euclidean spatial geometries
$D(o,e)$ is taken to be the angular-diameter
distance in the background geometry from the observer
to the intersection of the $x$-axis with the source plane,
and $D(l,e)$ is taken to be the angular-diameter distance in the
background geometry
between the deflection point, $p$, and the intersection
of the $x$-axis with the source plane, $p^{\prime}$.
Then, again assuming small deflection angle,

$$\beta^i =\theta^i -{D(l,e)\over D(o,e)}\alpha^i .         \eqno{(1)}$$

\noindent
This is the standard cosmological gravitational lens equation
(Schneider, Ehlers, and Falco 1993).
The magnification matrix, $M^i{}_j=
\partial \beta^i /\partial\theta^j $
contains information about the deformation of ray bundles
connecting the observer and emitter.
For example, the inverse of the determinant of this
matrix is the magnification of an image
relative to an unlensed image.

The purpose of the current paper is to address
a number of subtle issues
that arise when we attempt to justify
mathematically the use of the
lens equation (1) for calculations in our Universe, although most workers
agree that the physical justifications for its use in observed lens
systems are strong.
First, there is the question of the best choice of distance factors.
There exists a large
literature addressing this question,
primarily concerned with the appropriateness
of the so-called Dyer-Roeder distances
(Dyer and Roeder 1972, 1973; Ehlers and Schneider 1986;
Futamase and Sasaki 1989; Watanabe and Tomita 1990; Watanabe, Sasaki,
and Tomita 1992; Sasaki 1993).
Our work suggests that within the framework of
cosmological perturbation theory, the natural
distance factors to use are those of the background.
Hence, the choice of distance factors
is equivalent to the choice of cosmological
model, in agreement with the recent results
of Sasaki (1993). The issue of which
cosmological model is most appropriate
must be addressed in its own right.

The second issue which should be addressed in
any mathematical investigation into the lens
equation concerns the accuracy of the approximation of
an actual photon path by
two geodesics of the background which join
at a point near the lens, the deflection point, $p$.
On physical grounds we expect this approximation
to be good for systems for which the photon-lens
interaction is localized: the thin-lens approximation.
One purpose of our present work is to quantify the relationship
between the actual path and that used in the lens equation.

Third, how are we to find the angles
appearing in the lens equation from physical data?
Generally, the $\alpha^i$ are taken to be
those predicted by calculations in
Einstein-de Sitter spacetime,
since the overall curvature of space should not be important
near $p$, where the light ray
interacts with the lensing object.
For static, thin lenses, these calculations
write the deflection angle as a superposition of
point mass deflection angles
contributed by each mass element of the lens projected onto the
lens plane (Schneider et al. 1993).
For brevity we will term the resultant angle
the ``Einstein angle.''
Another purpose of this work is to derive this result
from the full equations of light propagation under
an appropriate set of mathematical approximations.
The lens equation, for static, thin lenses,
effectively assumes that
the light path is described by the Jacobi equation of the
background spacetime subject to an impulsive
wavevector deflection at the lens
plane by an angle equal to the usual
Einstein bending angle. We wish to quantify the level of
approximation involved.

There have been two notable recent attempts to clarify
the validity of the cosmological
lens equation by deriving it
from the optical scalar equations
(Seitz, Schneider, and Ehlers 1994) and the Jacobi
equation (Sasaki 1993).
However, a crucial difference between these papers
and the present work is that they treat the path
of the light ray differently near to
and far from the lens.
It is precisely this
assumption that we must eliminate if we
hope to gain a more general lens equation able
to quantify the errors implicit in equation (1).

Our
approach is to investigate
the cosmological lens equation as it emerges from the geodesic
equation.
We are able to do this by making use of a
technique for constructing null geodesics in
perturbed spacetimes (Pyne and Birkinshaw 1993).
The conventions of the present paper are the same
as those in that earlier work.
The results of this letter come from applying this technique to
FRW spacetimes with scalar perturbations in the longitudinal gauge. The
calculations yielding the results presented here will
appear in Pyne and Birkinshaw (1994). The results we report here are:
analytic formulae (equations (4) and (5)) for light
rays in the spacetime (2); a general expression for the magnification
undergone by a bundle of light rays
capable of handling non-static,
geometrically thick, lenses (equations (10), (11), and (13));
and a demonstration that the usual deflection angle, lens
equation (1) and magnification
matrix are recovered in the appropriate approximations.
To our knowledge, these are the first rigorous demonstrations
of these results for perturbed FRW spacetimes.

\noindent
{\bf 2. The Deflection Angle}

Our starting point is a choice for the metric of our Universe.
We choose to work
with FRW spacetimes with scalar perturbations
written in the longitudinal gauge,

$$d{\bar s}^2=a^2\left[-(1+2\phi )d\eta^2 + (1-2\phi )
\gamma^{-2}\left( dx^2+dy^2+dz^2 \right)\right]  \eqno{(2)}  	$$

\noindent
where
$\gamma =1+\kappa r^2/4$, $\kappa$ being the spatial curvature
parameter
($\pm 1$ or 0) and
$r^2=x^2+y^2+z^2$. Inhomogeneities are represented by the
quasi-Newtonian potential, $\phi$. For the order needed by us the expansion
factor, $a$, is unperturbed from its Friedmann form
(Jacobs, Linder, and Wagoner 1993).
The metric (2) is also used by
Seitz et al. (1994) and Sasaki (1993),
so that direct comparison of results is possible, and
recent work by Futamase (1989) and Jacobs et al.
(1993) has shown that
structure of galactic scales
and greater in our Universe can be
well modeled by metrics of this type.
While results obtained with the metric (2) are not appropriate
for lensing by gravitational waves or vector perturbations,
it is possible to
treat these cases by a similar method
(see Pyne and Birkinshaw 1994).

Because the Friedmann expansion, $a$, plays the role of a conformal
factor  it is simplest to work with the null geodesics of
$ds^2$, defined by $d{\bar s}^2=a^2ds^2$.
Quantities in $d{\bar s}^2$ will always be written
with an overbar.
Light rays in these two metrics coincide and their
(affine) parameterizations are related by ${\bar
k}^{\mu}=a^{-2}k^{\mu}$.
While our formulae are mostly written in terms of
quantities in $ds^2$, our results apply to the actual spacetime,
$d{\bar s}^2$. We note that by locating the observer at the spatial origin
the radial null geodesics of $ds^{(0)2}$,
(i.e. of that part of $ds^2$ independent of $\phi$),
obey
$k^{(0)0}=1$ and $k^{(0)i}=
-\gamma e^i$, where $e^i$ are the direction cosines at the observer,
so that $\sum_{i=1}^{3}(e^i)^2=1$ (McVittie 1964).
The explicit solutions for the comoving radius and for $\gamma$ along
such rays are given by

$$\eqalign{r (\lambda )    &=2\tan_{\kappa }\left(
{\lambda_o-\lambda \over 2}\right)  \cr
          \gamma (\lambda ) &=\sec_{\kappa }^2\left(
{\lambda_o-\lambda \over 2}\right) \cr} \eqno{(3)} $$

\noindent
where $\lambda_o$ is the affine
parameter at the observer. The subscript $\kappa$ on a
trigonometric function denotes a set of three functions:
for $\kappa =1$ the trigonometric function itself,
for $\kappa =-1$ the corresponding hyperbolic
function, and for $\kappa =0$ the first term
in the series expansion of the function.
The paths of the rays
are $x^{(0)0}=\lambda$, $x^{(0)i}=re^i$.

The method presented in Pyne and Birkinshaw (1993) allows
us to write the null geodesics of $ds^2$
as $x^\mu=x^{(0)\mu}+x^{(1)\mu}$.
For light rays passing through the observer
$x^{(0)\mu}$ is given as above and
the separation vector is

$$x^{(1)0}(\lambda )=+2\int_{\lambda_o}^{\lambda}\, du
(u-\lambda)k^{(0)m} \phi_{,m}(u)
			\eqno{(4)}$$

\noindent
and

$$\eqalign{x^{(1)i}(\lambda )
			&=-2k^{(0)i}\int_{\lambda_o}^{\lambda}\,
du (u-\lambda){\partial \phi\over \partial \eta}(u)	\cr
			&\qquad +2\gamma (\lambda )
\int_{\lambda_o}^{\lambda}\, du
\sin_{\kappa } (u-\lambda){1\over \gamma (u)}
\left( \nabla_{\perp }\phi \right)^i\cr}
					\eqno{(5)}$$

\noindent
where $\nabla_{\perp }{}^i=g^{(0)mn}\left(
\delta^i_n -e_ne^i\right)\partial_m $ is the transverse
gradient operator. We see that the spatial
separation is naturally written as a the
sum of a longitudinal and
a transverse term. The above integrals are taken over
the background geodesic $x^{(0)\mu}$.
Hence if these solutions are used for
geometrically thick lenses, it will in general
be necessary to apply an iterative procedure,
incorporating a number of background paths. Such an
approach is familiar from the usual multiple-lens
plane theory (Schneider {\it et. al.} 1993)
but we stress that formulae (4) and (5)
allow the photon path be approximated
to arbitrary accuracy, in contrast with the
multiple lens plane theory where the continuum limit
is not compatible with the asumptions underlying
the theory. Precise statements
of consistency will appear in Pyne and Birkinshaw (1994).

We can understand (4) and (5) by considering
their relation to the equation of geodesic deviation.
The Jacobi equation of $ds^{(0)2}$ subject to an
arbitrary impulsive wavevector perturbation $\delta k^{\mu}$ at some
affine parameter $u$, is solved by deviation
vector $\delta x^{\mu}$
with spatial components

$$\delta x^i(\lambda )=-{\gamma (\lambda )
\over \gamma (u)}\sin_{\kappa } (u-\lambda )
\delta k^i_{\perp}(u)-{\gamma (\lambda )\over \gamma (u)}
(u-\lambda )\delta k^i_{\parallel}(u)          \eqno{(6)}$$

\noindent
where $\delta k^i_{\perp}=\left( \delta^i_j-e^ie_j\right)\delta k^j$
is the impulse in the transverse direction,
and $\delta k^i_{\parallel}=
\delta k^i-\delta k^i_{\perp}$ is the longitudinal impulse.

Comparing (5) with (6), produces the following interpretation of
the spatial components
of our solution for the separation, $x^{(1)i}$:
the photon path differs from a background path
because of a continuous sequence of impulsive perturbations

$$\delta k^i =-2\left( \nabla_{\perp }\phi \right)^i
 +2k^{(0)i}{\partial \phi \over \partial \eta} .\eqno{(7)}$$

\noindent
In fact, the form of the impulse can be gained directly
from our equations by
differentiating the spatial separations with
respect to the affine parameter, and inserting a delta
function at $\lambda_l$ into the integrand which forces the
integrand to vanish except at the lens plane.
This gives

$$ k^{(1)i}\left(\lambda_l\right) =-2\left(
\nabla_{\perp}\phi\right)^i \left( \lambda_l\right)
+2k^{(0)i}\left( \lambda_l\right)
{\partial \phi \over \partial \eta}\left(
\lambda_l\right), 	\eqno{(8)}$$

\noindent
which
is exactly the impulse found above, equation (7).

At this point we need only do a little work to
recover the Einstein angle from our equations.
Establish a set of cartesian axes at the observer
and choose the unperturbed wavevector $k^{(0)}=(1, -\gamma ,0,0)$.
Consider a static, localized perturbation in the
$xy$-plane.
Then the
angle represented by the impulse perturbation
found above, that is, the angle $k^{(0)i}\left(
\lambda_l\right)+k^{(1)i}\left(
\lambda_l\right)$ makes with $k^{(0)i}
\left( \lambda_l\right)$, is given by

$${\hat \alpha}^y={-2\left( \nabla_{\perp}\phi \right)^y\over
\gamma \left( \lambda_l\right)}=-2\gamma \left( \lambda_l
\right) \phi_{,y}.  \eqno{(9)}      $$

\noindent
The factor of $\gamma \left(\lambda_l \right)$
is present only because our co-ordinates are
scaled in an unusual
way at the lens plane. If we make the co-ordinate change $x^{i\prime}=
x^i/\gamma\left( \lambda_l \right)$
the metric on the lens plane becomes
Minkowskian and, locally near
the deflector,
$y'$ and $z'$ serve as normal co-ordinates on the lens plane.
In these co-ordinates
$\phi_{,y^{\prime}}$ on the lens plane takes on the usual Newtonian
form (with origin shifted from the lens, accounting for the
unusual minus sign).
Since our lensing angle $-2\gamma\left(
\lambda_l\right)\phi_{,y}=-2\phi_{,y^{\prime}}$
the integrated impulse
lensing angle for a localized perturbation
is exactly the Einstein deflection angle.
We emphasize that this is the first time this
has been shown rigorously for the curved FRW spacetimes.

For completeness we note that
the timelike component of the separation can also be analyzed
by comparison to the Jacobi equation.
The Jacobi equation of $ds^{(0)2}$ for
an impulse wavevector perturbation $\delta k^{\mu}$ at affine parameter
$u$ results in a timelike component of the deviation vector
$\delta x^0=-(u-\lambda )\delta k^0(u )$. Comparison
with our solution for the separation reveals that the time
delay undergone by the light ray relative to the fiducial background
ray may be considered to result from a sequence of impulses
$\delta k^0=-2k^{(0)m}\phi_{,m}$.

\noindent
{\bf 3. The Magnification}

We now want to examine the magnification undergone by a bundle of light rays.
We define this after Schneider et al. (1993)
in the following way.
Suppose a source of given physical size at some redshift
is observed to subtend solid angle $d\Omega$. An identical
source observed at identical
redshift placed in an FRW spacetime would
subtend solid angle $d\Omega^{(0)}$.
The magnification $M$ is defined to
be $d\Omega / d\Omega^{(0)}$.

We can construct a bundle of light rays in the spacetime (2)
which emanate from a source and converge at an observer by varying
the direction cosines, $e^i$, in equations (4) and (5).
This yields $d\Omega$ in terms of the redshift and proper size
of the source. A similar relationship is easy to
derive for a source in the background spacetime. Comparison of the
two yields $M$. If the four-velocities of the observer
and emitter are written
$u_{o(e)}^{\mu}=(1-\phi_{o(e)}, v^i_{o(e)})$, which are correct
to first order in $\phi$ and $v$, then

$$ \eqalign{ M	&=\left( 1+2\phi_o+2
\left[ v^ik^{(0)}_i\right]^e_o-2k^{(1)0}_e
+2\cot_{\kappa} \left(\lambda_o-\lambda_e\right)\delta\lambda_e+
\kappa\sin_{\kappa }\left(
\lambda_o-\lambda_e\right)e_ix^{(1)i}_e\right) \cr
	&\qquad /\left({\rm Det}M^i{}_j \right)\cr} \eqno{(10)}$$

\noindent
where $\left[ f\right]^e_o=f\left( \lambda_e\right)
-f\left(\lambda_o\right)$ and the subscript $e$
indicates evaluation at $\lambda_e$.
Here $\delta\lambda_e$ enforces the equal redshift aspect of
the definition. It
is set by
$1+z(\lambda_e+\delta\lambda_e )
=1+z^{(0)}(\lambda_e)$
and is given by

$$\delta\lambda_e={a_e\over {\dot a}_e}\left[ \phi-v^ik^{(0)}_i
-{\dot a}a^{-1}x^{(1)0}+k^{(1)0}\right]^e_o    . \eqno{(11)}$$

\noindent
$M^i{}_j$ is the magnification matrix, of dimension $2\times 2$,

$$M^i{}_j=I_2+{1\over r}{\partial x^{(1)i}\over \partial e^j} ,
\eqno{(12)}$$

\noindent
where $I_2$ is the two-dimensional identity matrix and
the indices $i,j$ run only over the transverse spatial
dimensions. For simplicity, we will henceforth assume the
unperturbed wavevector $k^{(0)}=(-1, \gamma ,0,0)$ so that
$i,j$ run over
$\lbrace 2,3\rbrace $. Keeping in mind the transverse nature
of these indices we can write
$\partial /\partial e^j=\partial /\partial\theta^j $, with
${\bf \theta}$ the vectorial angle of equation (1).

Explicit calculation yields

$$\eqalign{ M^i{}_j=\delta^i_j  &
+{2\over \sin_{\kappa}\left( \lambda_o-\lambda_e\right)}
\delta^i_j\int_{\lambda_o}^{\lambda_e}du\, \left(u-\lambda _e\right)
{\partial \phi
\over \partial\eta}(u) \cr
                        &-{2\over \sin_{\kappa}
\left( \lambda_o-\lambda_e\right)}
\int_{\lambda_o}^{\lambda_e}du\, \sin_{\kappa}\left( u-\lambda_e\right)\gamma
(u) \cr
	&\times \left[ \delta^i_j{\partial \phi\over \partial x}(u)
-r(u)\gamma^{-2}(u)g^{(0)ik}(u)
\phi_{,kj}(u)\right]. \cr }\eqno{(13)}$$

\noindent
To see how the usual magnification matrix is
contained in equation (13) we rewrite
the second integral
using the angular diameter distance of $d{\bar s}^{(0)}$,
${\bar D}(u,\lambda )=a(\lambda )
\sin_{\kappa} (u-\lambda)$. In this way

$$\eqalign{ M^i{}_j=\delta^i_j	&+{2\over \sin_{\kappa }
\left( \lambda_o-
\lambda_e \right)}\delta^i_j\int_{\lambda_o}^{\lambda_e}du\,
\left( u-\lambda_e\right){\partial \phi\over \partial \eta}(u) \cr
	&-{1\over {\bar D}\left( \lambda_o,\lambda_e\right)}
\int_{\lambda_o}^{\lambda_e}du\,
{\bar D}\left( u,\lambda_e\right)
{\partial {\hat \alpha}^i\over \partial \theta^j}, \cr}   \eqno{(14)}$$

\noindent
with the two-dimensional projected angle

$${\hat \alpha}^i=-2{\left( \nabla_{\perp}
\phi\right)^i\over \gamma (u)}    . \eqno{(15)}$$

\noindent
We now consider lensing by a static, geometrically thin lens. The
first term in equation (14) vanishes and we approximate the angular
diameter factor as constant over the region for which the
potential is important so that

$$M^i{}_j\approx \delta^i_j-{{\bar D}
\left( \lambda_l,e\right)\over {\bar D}(o,e)}
{\partial \over \partial\theta^j}
\int_{\lambda_o}^{\lambda_e}du\,
{\hat \alpha}^i     \eqno{(16)}$$

\noindent
We have already seen that the integral of ${\hat \alpha}^i$ over the background
path
produces the Einstein deflection angle.
As a result we conclude that our equation
has reproduced the usual magnification matrix defined
by the $\theta$-gradient of equation (1).

\noindent
4. {\bf Summary}

We have presented formulae (4), (5) for the null geodesics
intersecting an observer's worldline in an important class of
perturbed spacetimes, FRW backgrounds with scalar
perturbations, in the longitudinal gauge. We have used these equations
to obtain a general formula (10) for the magnification of ray bundles in
these spacetimes. With this, we can show
for the first time how the usual lens equation (1) and
magnification matrix are recovered in these spacetimes without
dividing light paths into near and far lens regions.
In forthcoming papers we will consider
the implications of these results for practical
lensing calculations.

\vskip0.5true in

\noindent
{\bf Acknowledgements}

We thank Ramesh Narayan, Sylvanie Wallington,
and especially Sean Carroll for many helpful discussions.
This work was
supported by the National Science Foundation under grant AST90-05038.

\vskip1.0truein
\noindent
{\bf References}

\vskip 12pt
\normalbaselineskip=8pt plus0pt minus0pt
                            \parskip 0pt

\def\ref#1  {\noindent \hangindent=24.0pt \hangafter=1 {#1} \par}
\def\vol#1  {{\bf {#1}{\rm,}\ }}
\ref{Dyer, C. C., \& Roeder, R. C. 1972, ApJ,
174, L115}
\ref{Dyer, C. C., \& Roeder, R. C. 1973, ApJ,
180, L31}
\ref{Ehlers, J., \& Schneider, P. 1986, A\&A,
168, 57}
\ref{Futamase, T. 1989, MNRAS, 237, 187}
\ref{Futamase, T., \& Sasaki, M. 1989, Phys. Rev.
D., 40, 2502}
\ref{Jacobs, M. W., Linder, E. V., \& Wagoner, R.~V. 1992,
Phys. Rev. D., 45, R3292}
\ref{McVittie, G. C. 1964, General Relativity and Cosmology
(2d ed., London: Chapman and Hall}
\ref{Pyne, T. \& Birkinshaw, M. 1993, ApJ, 415, 459}
\ref{Pyne, T., \& Birkinshaw, M. 1994, in preparation}
\ref{Sasaki, M. 1993, preprint}
\ref{Schneider, P., Ehlers, J., \& Falco, E.~E. 1993,
Gravitational Lensing (Berlin: Springer-Verlag)}
\ref{Seitz, S., Schneider, P., \& Ehlers, J. 1994, preprint}
\ref{Watanabe, K., Sasaki, M., \& Tomita, K. 1992,
ApJ, 394, 38}
\ref{Watanabe, K., \& Tomita, K. 1990, ApJ, 355, 1}

\normalbaselineskip=24pt plus0pt minus0pt
                  \parskip 12.0pt
\vfill\eject

\noindent
{\bf Figure Captions}

\noindent
Figure 1. Gravitational lensing by a single symmetric lens.

\bye